\documentclass[prx,twocolumn,aps,epsf,showpacs,superscriptaddress,longbibliography,nofootinbib]{revtex4-2}

\usepackage[pdftex]{graphicx}
\usepackage{dcolumn}
\usepackage{bm}
\usepackage{epsfig}
\usepackage{latexsym} 
\usepackage{amsmath}
\usepackage{amssymb}
\usepackage{color}
\usepackage{array}
\usepackage{bbm}
\usepackage{hyperref}
\usepackage{color}
\usepackage{array}
\usepackage{cancel}
\usepackage{amsfonts}
\usepackage{verbatim}
\usepackage[dvipsnames]{xcolor}
\usepackage{ulem}

\usepackage{titlesec}
\titleformat{\paragraph}[runin]
        {\bfseries}
        {}
        {0.0em}
        {}
        [~ -- ~]
\titlespacing*{\paragraph}{0pt}{4pt}{0pt}

\newcommand{\<}{\langle}

\newcommand{\up}{\uparrow}
\newcommand{\down}{\downarrow}

\renewcommand{\>}{\rangle}
\renewcommand{\(}{\left(}
\renewcommand{\)}{\right)}
\renewcommand{\[}{\left[}
\renewcommand{\]}{\right]}
\renewcommand{\v}[1]{\boldsymbol{#1}} 
\renewcommand{\vec}[1]{\v{1}}

\newcommand{\bs}[1]{\boldsymbol{#1}}
\renewcommand{\d}{\partial}

\renewcommand{\L}{\mathcal{L}}

\newcommand{\vpd}{\vphantom\dagger}

\newcommand{\moire}{moir\'e }

\begin{document}
\title{Gate-tunable heavy fermion quantum criticality in a moir\'e Kondo lattice}
\author{Ajesh Kumar}\thanks{These authors contributed equally to this work.}
\affiliation{Department of Physics, University of Texas at Austin, Austin, TX 78712, USA}
\author{Nai Chao Hu}\thanks{These authors contributed equally to this work.}
\affiliation{Department of Physics, University of Texas at Austin, Austin, TX 78712, USA}
\author{Allan H. MacDonald}
\affiliation{Department of Physics, University of Texas at Austin, Austin, TX 78712, USA}
\author{Andrew C. Potter}
\affiliation{Department of Physics, University of Texas at Austin, Austin, TX 78712, USA}
\affiliation{Department of Physics, University of British Columbia, Vancouver, Canada}

\begin{abstract}
We propose a realization of Kondo-lattice physics in \moire superlattices at the interface between a WX$_2$ homobilayer and MoX$_2$ monolayer (where X=S,Se). Under appropriate gating conditions, the interface-WX$_2$-layer forms a triangular lattice of local moments that couple to itinerant electrons in the other WX$_2$-layer via a gate-tunable Kondo exchange interaction. Using a parton mean-field approach we identify a range of twist-angles which support a gate-tuned quantum phase transition between a heavy-fermion liquid with large anomalous Hall conductance and a fractionalized chiral spin-liquid coexisting with a light Fermi liquid, and describe experimental signatures to distinguish among competing theoretical scenarios.
\end{abstract}
\maketitle

\begin{figure*}[t]
\centering
\includegraphics[width=\textwidth]{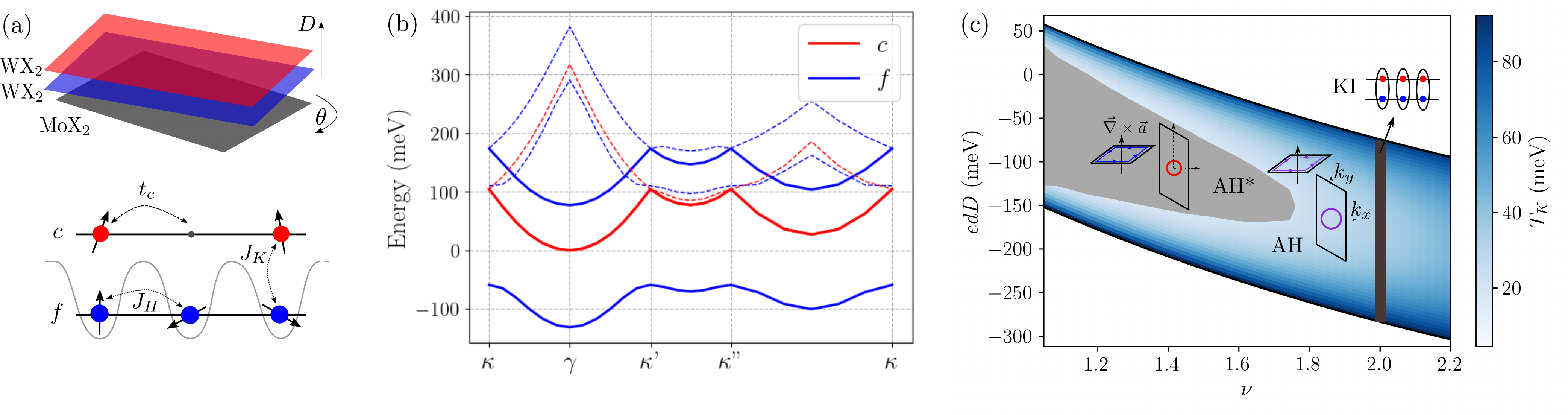}
\caption{\textbf{Kondo Lattice Simulator:}
(a) Schematic of a TMD trilayer Kondo lattice system.  The two active TMD layers experience moir\'e modulation 
potentials with different strengths so that one layer has flat bands and one has dispersive bands.  
When the density in the flat band layer is tuned to one electron per moir\'e period, and the total density to a higher value,
a triangular lattice Kondo model is realized. (b) Gated self-consistent Hartree-Fock band structures in the middle of the Kondo lattice model regime at $\theta=$ 4.5$^{\circ}$, with filling factors $\nu_f = 1$ and $\nu_c = 0$. 
The solid blue bands are the upper and lower Hubbard bands in the f-layer. The dashed blue bands are the second lowest energy $f$-band in each spin. The red bands are the first (solid) and second (dashed) lowest energy (spin-degenerate) bands in the c-layer with bandwidth $104.49$meV. (c) Parton mean-field phase diagram as a function of the 
moir\'e band filling factor $\nu$ and the external displacement field $D$, which are linearly related to the top and 
bottom gate voltages.  The black lines bound the Kondo lattice regime within-which the mean-field filling factor is $1$ in the 
f-layer and $\nu_c=\nu-1$ in the c-layer.  In the AH* phase, a Fermi surface of $c$-electrons (red) 
is decoupled from a chiral spin liquid (CSL) of spinons $f$. In the AH phase the Fermi surface (purple) has 
mixed $c$ and $f$ character.  The Kondo insulator (KI) phase at $\nu=2$ is adiabatically connected to 
direct product of singlets between $c$ and $f$ electrons on each site.  
The color scale indicates the Kondo temperature of the AH states.}
\label{fig:setup} 
\end{figure*}

Crystals containing lanthanide or actinide elements host $f$-electron local spin moments coupled via Kondo spin-exchange interactions to itinerant conduction ($c$) electrons. The phase diagrams of such Kondo lattice systems are often extremely rich, and
can include magnetic and non-magnetic states, superconductors, and non-Fermi liquids 
~\cite{stewart1984heavy,gegenwart2008quantum,paschen2020quantum,coleman2001fermi,senthil2004weak}.
When Kondo screening effects dominate, the $f$-spins hybridize with $c$-electrons and contribute to the Luttinger-volume of the Fermi-sea producing exceptionally high (heavy fermion) quasiparticle masses~\cite{stewart1984heavy,gegenwart2008quantum,paschen2020quantum} arising from the narrow $f$ bandwidth. Alternatively, dominant spin-spin interactions (chiefly c-mediated by RKKY coupling) cause $f$ local moments to order into either i) a magnetic state~\cite{doniach1977kondo}, which is typically an antiferromagnet (AFM)
or ii) a quantum spin liquid (QSL)~\cite{coleman1989kondo,senthil2004weak} if magnetic order is frustrated. 

These Kondo-(un)screened phases are separated by a quantum critical point marked by a fan of non-Fermi liquid transport signatures~\cite{gegenwart2008quantum,paschen2020quantum,coleman2001fermi,lohneysen2007fermi} that is often partly occluded by an unconventional superconducting dome. The hallmark of this apparently-continuous critical point is a sudden change in the size of the Fermi surface. Understanding the nature of such Fermi-volume-changing transitions and their relation to non-Fermi liquid behavior is thought to be key to understanding many other strongly correlated systems such as high-temperature cuprate superconductors, and remains a key unsolved problem in condensed-matter physics. Moreover, different experiments and theoretical analyses produce a conflicting picture of the nature of the non-hybridized phase, and its nature remains hotly debated despite decades of intensive study. The complex microscopic structure, many-band character of $f$-electron materials along with the limited ability to tune carrier-densities and exchanging couplings in bulk $3d$ crystals pose significant obstacles to resolving these mysteries.

In this Letter we propose that progress can be achieved by using recently discovered 
transition metal dichalcogenide (TMD) moir\'e material platforms to construct a synthetic Kondo lattice made from relatively simple well-understood components. Crucially, we will show that the Kondo-exchange coupling can be continuously tuned by electrostatic gate tuning, enabling in-situ access to the entire heavy-fermion phase diagram and criticality in a single device. Furthermore, these systems have triangular lattice symmetries that tend to frustrate magnetic order, and can favor fractionalized QSL states.  Our main results are summarized in Fig.~\ref{fig:setup} which contains phase diagrams in the two-dimensional gate-voltage phase space of dual-gated two-dimensional materials.
We first identify the twist-angle dependent area within which a Kondo lattice model is realized and then  estimate the location of the Kondo-screening quantum critical line. Parton mean-field calculations predict 
a novel quantum criticality scenario in which local moments in a weak-coupling chiral spin liquid (CSL) state hybridize
with conduction electrons to form a strong-coupling heavy Fermi liquid that exhibits an anomalous Hall effect.  
We discuss experimental probes of Kondo lattice physics in moir\'e materials
that discriminate among various competing antiferromagnetic (AFM) and quantum spin-liquid (QSL) weak Kondo-coupling scenarios, highlighting electrical and thermal Hall transport, and electrostatic measurements of entropy.
Compared to previous proposals for achieving heavy fermion physics in \moire systems, our setup directly implements a Kondo lattice model using already-demonstrated \moire ingredients; it avoids both topological obstacles~\cite{zou2018band,po2019faithful,song2019all,ahn2019failure} to local-moment physics of proposed twisted graphene multilayer realizations~\cite{ramires2021emulating}, as well as the need for unusual spontaneous orbital selectivity required to form local moments in electron-doped TMD heterobilayers~\cite{dalal2021orbitally}.

\paragraph{A synthetic Kondo-Lattice}
We propose a realization of Kondo lattice physics in TMD
trilayers in which a triangular lattice moir\'e pattern is formed between an atomically aligned WX$_2$ bilayer twisted by a small angle $\theta$ relative to 
 an MoX$_2$ monolayer, where X=S,Se is a chalcogen. We note that one could alternatively form a \moire potential by choosing different chalcogens for the WX$_2$ and MX$_2$ layers to force a lattice mismatch (though we do not model that case here).
The active valence-band degrees of freedom reside in the WX$_2$ layers and come from momenta near the $K$ and $K'$ points in the 
triangular-lattice Brillouin zone, where strong spin-orbit coupling locks the spin and valley degrees of freedom into a single effective
spin-1/2 moment (with approximate SU(2)-symmetric interactions~\cite{xiao2012coupled,wu2018hubbard}) that we will refer to simply as spin. 
Because the bottom WX$_2$ layer (closest to the MoX$_2$) experiences a stronger moir\'e  
modulation potential \cite{wu2018hubbard}, its triangular-lattice symmetric moir\'e minibands will be narrower, enhancing correlations in this layer. Temporarily neglecting interlayer tunneling, the goal will be to half-fill the lowest moir\'e miniband of the lower (f) layer to form a Mott insulator of well-formed local moments, and partially fill the upper layer (c) to form a conducting Fermi surface.
Mott insulators in related TMD bilayers have been achieved in several recent experiments \cite{tang2020tTMD,regan2020mott,wang2020correlated,xu2020correlated,jin2021stripe,li2021charge,li2021imaging,fractionTMD2020,mak2021continuousMIT,dean2021quantumcritical,li2021quantum}, indicating the feasibility of this setup.  
Next, one must adjust the top-gate electric field $E_t$ so that  hole chemical potential, $\mu_c$, lies inside the charge gap of the f-layer, $0<\mu_c<U$ 
where $\mu_c = \d \epsilon_c/\d n_c$ is the chemical potential and $\epsilon_c$ is the energy per area of the c-layer,
$U$ is the charge gap of the f-layer, and we measure energy relative to the bottom of the f-layer gap.
If correlations are neglected in the $c$-layer 
\begin{equation}
    \mu_c = \Big(\frac{2 \pi e^2 }{\epsilon A_M} (\nu_c-1) +  e D\Big) \;d  + \epsilon_F(\nu_c) + U,
\end{equation}
where the displacement field $D \equiv (E_t+E_b)/2$, $\epsilon$ is the background dielectric constant,
$E_{(t,b)}$ are the gate electric fields above the c-layer and below the f-layer respectively, 
$d$ is the separation between WX$_2$ layers,
$\nu_c$ is the number of holes per moir\'e period in the top layer, and $\epsilon_F(\nu_c)$ is the 
c-layer Fermi energy. 
Since the total carrier density is related only to the difference in gate fields via the Poisson equation: $4\pi e(1+\nu_c)/(\epsilon A_M) = E_t-E_b$, $\nu_c$ and $D$ can be controlled separately. 
For a given value of $\nu_c$, $\mu_c$ increases monotonically with $D$ and passes through the $(0,U)$ Kondo-lattice interval.
As illustrated in Fig.~\ref{fig:setup}(c), $\mu_c(D)$ increases with $\nu_c$ at fixed $D$ 
and both boundaries of the Kondo lattice interval move to smaller $D$.

\paragraph{Kondo Lattice Model}
Since the on-site repulsion in the f-layer, $U$, greatly exceeds the f-layer hopping and f/c interlayer tunneling amplitudes, within the Kondo lattice regime highlighted in Fig.~\ref{fig:setup}c, we model the system by a Kondo-lattice Hamiltonian that retains only spin degrees of freedom in the f-layer:
\begin{align}
    H_\text{KH} =& -t_c\sum_{\<\v{r}\v{r'}\>} c^{\dagger}_{\v r \alpha} c^{\vpd}_{\v r' \alpha} + \frac{J_K}{2} \sum_{\v{r}} \v{S}_{\v r} \cdot c^{\dagger}_{\v{r} \alpha} \v{\sigma}_{\alpha \beta} c^{\vphantom\dagger}_{\v{r} \beta} \nonumber \\
    &- \mu_c \sum_{\v r}c^{\dagger}_{\v r \alpha} c^{\vpd}_{\v r \alpha}+ J_H \sum_{\<\v{r}\v{r'}\>} \v{S}_{\v{r}} \cdot \v{S}_{\v{r'}}.
    \label{eq:HKL}
\end{align}
 Here $\v{S}_{\v{r}}$ is a spin operator at a triangular lattice
site $\v{r}$ in the f-layer, $c^{\vphantom\dagger}_{\v{r} \alpha}$ annihilates a c-layer electron of spin $\alpha$ 
at site $\v{r}$ and the repeated spin labels are implicitly summed. 

We extract the conduction band hopping ($t_c$) and Kondo coupling ($J_K$) parameters 
from self-consistent Hartree-Fock (SCHF) calculations in which the small hybridization, 
$\Gamma$, between the WX$_2$ layers is temporarily neglected (see Appendix~\ref{app:schf} for details).
We later reintroduce $\Gamma$ perturbatively to compute spin-exchange couplings. 

Inside the Kondo lattice regime the solutions of the SCHF equations 
are characterized by a mean-field state that has a small itinerant electron Fermi surface, a fully occupied 
triangular-lattice~\cite{wu2018hubbard, tang2020tTMD,regan2020mott,wang2020correlated,hu2021}
majority-spin local moment band, and an empty minority-spin local moment band.
SCHF calculations are performed using a continuum moir\'e model for holes with moir\'e modulation
potentials in the two WX$_2$ layers that attract holes to lattice sites and are represented by 
a Fourier expansion $\sum_{\v b} V_m(\v b) \text{exp}(i \v{b} \cdot \v{r})$ involving only the first shell of reciprocal lattice vectors $\v{b}$~\cite{wu2018hubbard}. We use a moir\'e potential strength $V_m= 30$ meV for the 
f-layer, based on recent experimental estimates~\cite{shabani2021deep}, 
and a WSe\textsubscript{2} effective mass $m=0.35m_e$ \cite{wu2018hubbard}, where $m_e$ is the mass of an electron.
We see in Fig.\ref{fig:setup}(b) that the c-layer and the minority spin f-layer mean-field bands are 
nearly-free-electron-like, with very small band-gaps to higher mini-bands at the zone boundary. This arises due to screening effects that result in approximate cancellation between attraction by the moir\'e potential and repulsion from the f-layer holes. 
Despite the small energetic separation between the first and second c-layer mini-bands, in Appendix~\ref{app:schf} we show that tunneling between the $f$-orbital and the second $c$-band is suppressed, justifying the use of a single-band model for both layers.

The mean-field Hartree potential at each total moir\'e band filling factor depends on the gate-controlled 
external displacement field $D$. 
The phase boundaries in Fig.~\ref{fig:setup}(c), calculated for the twist angle 
$\theta=4.5^\circ$ case, were constructed by identifying the area in the gate-voltage phase space over which the top layer Fermi level 
lies in the bottom layer charge gap. Data for other twist angles is summarized in Appendix~\ref{app:schf}.

To estimate the Kondo coupling constant, we perturbatively reintroduce the weak interlayer hybridization, which we take to be momentum-independent (see Appendix~\ref{app:schf}), via a Schrieffer-Wolff transformation to obtain:
\begin{align}
J_K = 2\Gamma^2 \( \frac{1}{U+E^f_F-E^c_F} - \frac{1}{E^f_F-E^c_F} \),
\end{align}
where $E^{f/c}_F$ are the respective Fermi levels. Crucially, we see that $J_K$ can be enhanced with a displacement field that brings the $c$ Fermi level closer to resonance with the upper or lower Hubbard band of the f-layer. Below, we will show that, for suitable twist angles, this allows gate-tuning across a Kondo-screening quantum critical point (KS-QCP).

The Heisenberg exchange coupling, $J_H$ between the $f$-spins arises both through super-exchange $\sim 4t_f^2/U$ between the f-layer spins, and via RKKY interactions mediated by the conduction electrons~\cite{aritsov1997indirect}:
$J_\text{RKKY}(\v{r}) =(\nu/\sqrt{3}t_c) J_K^2\[ \mathcal{J}_0(k_Fr)\mathcal{Y}_0(k_Fr)+\mathcal{J}_1(k_Fr)\mathcal{Y}_1(k_Fr)\]$
where $k_F$ is the conduction electron Fermi momentum, $\mathcal{J}_n,\mathcal{Y}_n$ are respectively Bessel functions of the first and second kinds\footnote{To smoothly interpolate to the strong coupling regime, $J_\text{RKKY}\sim J_K$  when $J_K \gg t_c$, in calculations we divide this perturbative formula by a factor $\sqrt{1+(J_K/t_c)^2}$ which produces the correct strong- and weak- coupling asymptotics.}.

\paragraph{Gate-tuned Kondo physics}
The moir\'e Kondo lattice system has two especially attractive features:
i) it is possible to tune the system between the strong $J_K$ heavy fermion regime and the more complex weak $J_K$ regime electrostatically by moving through the 
phase diagram with gates, and 
ii) the intrinsic triangular lattice geometry frustrates magnetic order and favors the more interesting fractionalized 
states that are thought to be a strong possibility in the weak $J_K$ regime.  
To identify where these states are most likely to occur  
we employ a parton\footnote{sometimes indelicately referred to as ``slave-spin"} mean-field theory~\cite{hewson1997kondo,senthil2004weak}, in which we fractionalize the f-layer spins into neutral spinons $\v{S}_{\v{r}} =\frac{1}{2} f^\dagger_{\v{r}\alpha}\bs{\sigma}_{\alpha\beta}f^{\vphantom\dagger}_{\v{r}\beta}$ which introduces a local-$U(1)$ gauge redundancy $f^{\vpd}_{\v r\alpha}\rightarrow e^{i\chi_{\v r}}f^{\vpd}_{\v r\alpha}$. In an exact treatment this results in an emergent dynamical $U(1)$ gauge field, $a_\mu$, which is approximated as a static background field in the mean-field theory (fluctuations beyond mean-field are discussed below where important). 

In the parton framework Kondo screening is captured by a non-vanishing value of a (non-local) hybridization order-parameter $\Delta_{\v r} = \<c^{\dagger}_{\v{r}\alpha}f^{\vpd}_{\v{r}\alpha}\>$. For $|\Delta|>0$ the neutral $f$-spinons hybridize with the charged $c$-electrons to become ordinary charged electrons. 
When the average background gauge flux $a_{\mu}$ is zero,
the system then has both a small Fermi surface of $c$-electrons with weak $f$ character, and a large heavy Fermi-surface of $f$-electrons with weak $c$ character; we denote this heavy Fermi liquid phase as FL. By contrast when the hybridization vanishes, $\Delta=0$, the spinons 
can form a time-reversal invariant QSL with a neutral Fermi surface of spinon excitations 
that coexists with the small Fermi surface of c-electrons. We denote this phase as FL* following Ref.~\cite{senthil2004weak}, which shows how gauge fluctuations beyond mean-field give strong non-Fermi liquid corrections to observable properties of the FL* phase and KS-QCP.

When ring-exchange effects are small, as is the case for the twist angles we consider, the parton mean-field-theory has been shown to instead favor mean-field solutions~\cite{motrunich2005variational}
with $\pi$ emergent magnetic flux per unit cell, which divide unequally between up and down triangular
plaquettes as illustrated in Fig.~\ref{fig_app:flux_pat}, spontaneously breaking time-reversal and lattice translation symmetries.
The emergent-magnetic field gives the spinon bands a net Chern number $C=\pm 1$, to form a chiral spin liquid (CSL). We note that, in this scenario, gauge-fluctuations are topologically-gapped due to a Chern-Simons term for $a_\mu$ induced by the spinon Chern bands. This mean-field prediction is further supported by recent 2d-DMRG~\cite{szasz2020chiral} calculations, that show regions of CSL phase arise in triangular-lattice Hubbard models near the Mott transition. 
Without hybridization ($\Delta=0$), CSL of $f$-spins coexists with the $c$-electron Fermi surface which we refer to as an Anomalous Hall* (AH*) phase. Kondo hybridization ($|\Delta|>0$), transmits the Berry curvature of the CSL-spins to the hybridized electrons resulting in a time-reversal-symmetry broken metal with a non-zero (but not quantized) anomalous Hall (AH) conductance. 

The parton mean-field phase diagram is shown in Fig.~\ref{fig:setup}(c) (see also Appendix~\ref{app:parton_mf}) for twist angle $\theta=4.5^\circ$. In all regimes sufficiently away from $\nu=2$ [where a Kondo Insulator (KI) arises], AH(*) phases are favored over flux-free FL(*) phases. We have also checked for magnetic mean-field solutions with magnetization, $\<\v{S}_{\v{r}}\>\neq 0$, forming the $120^\circ$ pattern favored 
by the triangular lattice Heisenberg model, but did not find any regions where AFM order either coexists with- or supplants- the AH or AH* phases. Crucially, the KS-QCP (here between AH and AH* phases) is accessible by gate tuning.
We estimate that the Kondo temperature, the characteristic temperature near which Kondo screening emerges, as $T_K\gtrapprox 100K$ 
using $T_K = \Lambda e^{-1/(2J_K \rho_c)}$~\cite{doniach1977kondo} where $\Lambda$ and $\rho_c$ are respectively the $c$-electron bandwidth and the density of states at the Fermi-level.  

Similar gate-field phase diagrams that include Kondo-screening quantum criticality lines arise over a range of twist-angles: $3.5^\circ\lesssim\theta\lesssim 6^\circ$. For larger $\theta$, the f-layer is no longer a Mott insulator, and for smaller $\theta$, the Kondo coupling is large compared to $t_c$ so that only hybridized phases arise. We note that, if desired, the quantum criticality could potentially be accessed at smaller twist angles by suppressing the interlayer tunneling, $\Gamma$ between WX$_2$ layers, for example, by introducing a hexagonal boron nitride (hBN) spacer layer.

\begin{figure}[t]
\centering
\includegraphics[width=1\columnwidth]{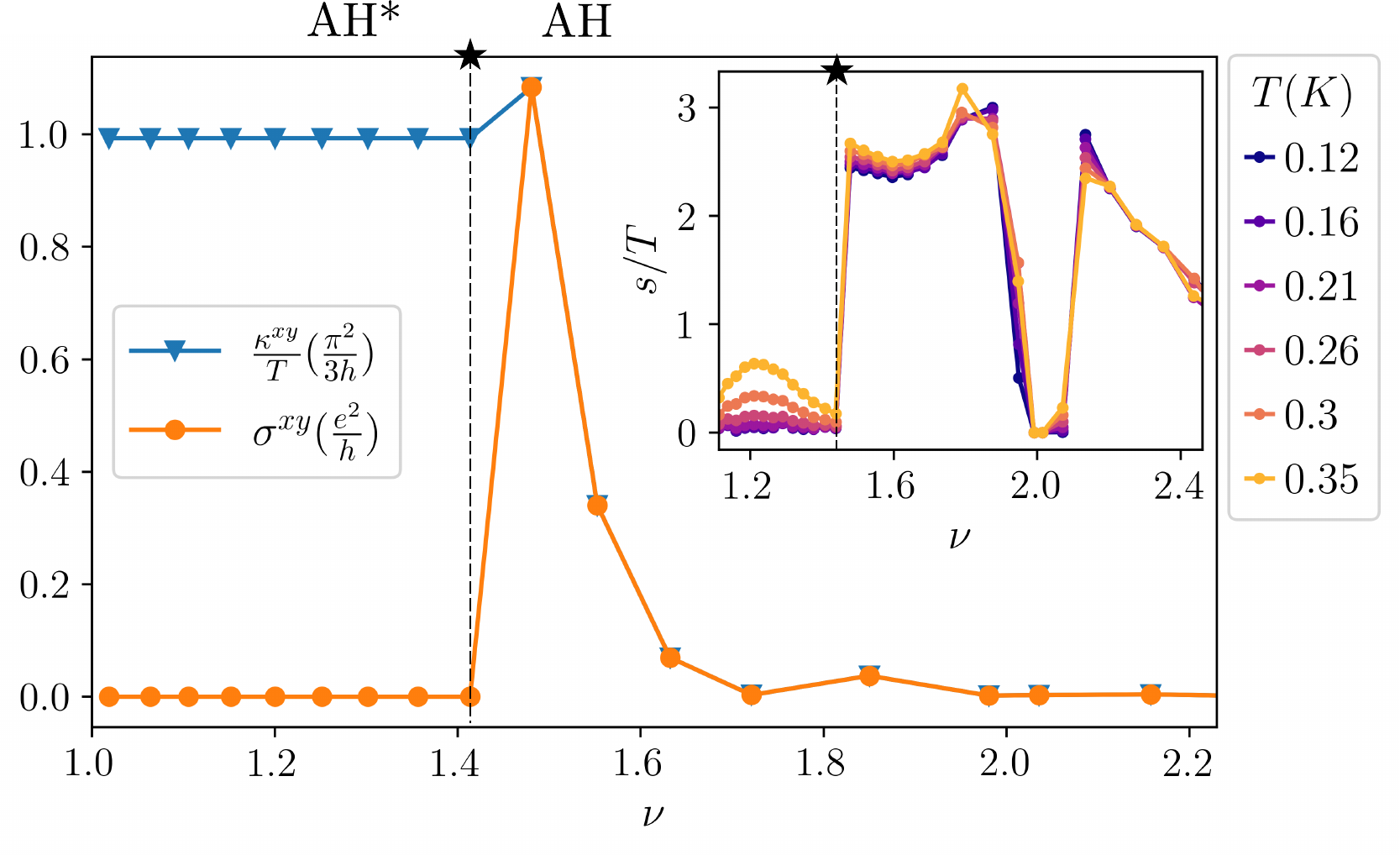}
\caption{Thermal ($\kappa^{xy}$) and electrical ($\sigma^{xy}$) Hall conductivities per spin, calculated using the parton mean-field bands for $\theta=4.5^{\circ}$ and $J_K=22.2$ meV. In the AH* phase, the spinons are in a Chern insulating state with quantized $\kappa^{xy}/T$ and a 
gap of $\approx 0.26$ meV at $\nu\approx 1$. $\sigma^{xy}$ jumps to the $\kappa^{xy}/T$ value in the AH phase as the spinons acquire electronic character. The inset shows $s/T$ per spin versus $\nu$ at temperatures specified by line color. 
The jump near the KS-QCP is due to effective mass enhancement in the AH phase when the hybridization exceeds
the Chern insulator gap.
 \label{fig:entropy} 
}
\end{figure}

\paragraph{Experimental signatures}
Existing measurement techniques for $2d$ moir\'e materials offer experimental tests that can both identify Kondo-screening quantum criticality line and 
cleanly distinguish among the various proposed QSL and AFM scenarios for the non-hybridized phase.
We focus on i) the entropy density $s(\nu)$, which can be extracted from electrostatic measurements of 
$\partial \mu/\partial T = -\partial s/\partial \nu$~\cite{kuntsevich2015strongly,hartman2018direct,rozen2021entropic,saito2021isospin}, 
and ii) electrical ($\sigma$) and thermal ($\kappa$) longitudinal and Hall conductivities, and present detailed parton calculations of these observables in Appendix~\ref{app:parton_mf}. Since thermal transport properties can be challenging to measure, in Appendix~\ref{app:interlayer_transport}, we also describe all-electrical probes of the spinon transport properties of QSL scenarios using a combination of AC measurements and separate contact of the c- and f- layers.

In each scenario, the $c$-electrons contribute Fermi-liquid like behavior with $s\sim T$ and $\sigma \sim \text{constant}$, that add to contributions of (neutral) $f$-spinons,with the exception of $\nu=2$ where we find a featureless KI with activated thermodynamic and transport signatures with gap $\approx 1.14$~meV. In each case, the Kondo-hybridization results in a discontinuous jump in electrical conductivities due to the incorporation of $f$-spins into the conducting Fermi-sea.

In the AH*/AH scenario favored by our parton calculations, the CSL of $f$-electrons in the AH* phase contribute a quantized thermal Hall conductivity $\kappa^{xy}_f=\frac{\pi^2}{3h}T$ per spin, and thermally-activated entropy $s\sim e^{-E_g/T}$ with $E_g$ the spinon-gap. Kondo hybridization produces a quantized jump $|\Delta\sigma^{xy}|=e^2/h$ per spin, but smoothly evolving $\kappa^{xy}$ across the KS-QCP. 
Alternatively, in a time-reversal symmetric FL*/FL scenario, the $f$-spinons form a gapless state with a Fermi-surface. Here, due to gauge-fluctuations beyond the parton mean-field treatment, the spinon entropy contribution varies from non-Fermi liquid-like $s\sim T^{2/3}$~\cite{lee1992gauge,senthil2004weak} in the FL* phase, marginal FL $s\sim T\log 1/T$~\cite{senthil2008theory} at the KS-QCP, and an ordinary $s\sim T$ in the FL phase. 
Lastly, in a more conventional AFM/FL scenario, the spinons are confined in the AFM phase where and the f-layer spin-waves would contribute $s\sim T^2$, as well as a jump in the $\sim T$ coefficient, Hall density, and conductivity across the KS-QCP. We note that the jump in electrical conductivities across the KS-QCP is expected to be much larger in the AFM/FL and FL*/FL scenarios than the AH*/AH scenario since the spinons are gapped at the KS-QCP in the latter.

 \paragraph{Discussion}
Several extensions of our theoretical analysis
could be fruitful avenues for future investigation.  Unfrustrated orthorhombic Kondo lattice systems, 
like the prototypical heavy fermion compound CeCoIn$_5$, often exhibit superconductivity in the vicinity of the KS-QCP. 
It remains an open question whether this superconducting tendency persists in frustrated triangular lattice geometries. 
Second, large moire unit cells imply large magnetic fluxes per unit cell~\cite{dean2013hofstadter}, and therefore significant 
scalar-chirality in local moment interactions, at relatively weak magnetic field $B$.  These could favor CSL phases~\cite{hu2016variational,wietek2017chiral} at finite $B$ even if they were not favored at
$B=0$.  The ability to apply large flux per unit cell could effectively induce orbital emergent 
gauge fields in the spinon Fermi surface state, leading to quantum oscillatory phenomena with $1/\kappa B$-periodicity, where $\kappa$ is an $O(1)$ gauge-susceptibility~\cite{motrunich2006orbital}. However, these effects may be challenging to observe in current TMD \moire samples, due to impurity suppression which so far obscures quantum oscillations even in simple Fermi-liquid states.
Another promising direction would be to explore the consequences of modifying the electronic structure by imposing relative twists between the two W-layers ~\cite{fengcheng2019homo, fengcheng2020DM, zang2021hartree}.

\noindent\vspace{4pt}{\it Acknowledgements -- } We thank Jie Shan and Qianhui Shi 
for insightful discussions. AK and AP were supported by the National Science Foundation through the Center for Dynamics and Control of Materials: an NSF MRSEC under Cooperative Agreement No. DMR-1720595. NCH and AHM were supported by the U.S. Department of Energy, Office of Science, Basic Energy Sciences, under Award \# DE‐SC0019481. AP acknowledges support from the Alfred P. Sloan
Foundation through a Sloan Research Fellowship.
This research was undertaken thanks, in part, to funding from the Max Planck-UBC-UTokyo Center for Quantum Materials and the Canada First Research Excellence Fund, Quantum Materials and Future Technologies Program.
Numerical calculations were performed using supercomputing resources at the Texas Advanced Computing Center (TACC).

\bibliography{moire_kondo}

\appendix
\onecolumngrid

\section{Hartree-Fock Calculations}\label{app:schf}
The single particle model of the system consists of two heterobilayer continuum models with different \moire potential strength are coupled by the real interlayer coupling $\Gamma$:
\begin{align}
    H = \begin{pmatrix}
    -\frac{k^2}{2m} + \sum_{i=1}^6 V_{mf}(\v{b}_i) \text{exp}(i \v{b}_i \cdot \v{r}) & -\Gamma\\
    -\Gamma & -\frac{k^2}{2m} + \sum_{i=1}^6 V_{mc}(\v{b}_i) \text{exp}(i \v{b}_i \cdot \v{r})
    \end{pmatrix},
\end{align}
where $V_m (\v{b}_j) = |V_m|\exp[(-1)^{j-1}i\phi]$. We fix $\phi=-94^{\circ}$ for all calculations, which means the lattice is triangular. The \moire potential strength is weaker for c-layer due to the larger distance to the MoX$_2$ layer. We model this effect by a simple exponential decay of the Coulomb interactions: $|V_{mc}| = |V_{mf}|\exp(-|\v{b}|d)$, where $d \approx 0.7~{\rm nm}$\cite{xiao2014effects} is the interlayer distance. 
Since $\Gamma$ is small compared to the energy difference between the diagonal terms in this Hamiltonian we neglect its effects in the Hartree-Fock calculations and model it later based on SCHF wave functions. The continuum model is then reduced to two decoupled single-band models, the energy cutoff and convergence properties of which largely follow twisted TMD heterobilayers \cite{hu2021}. The interaction-renormalized itinerant bandwidth has a strong dependence on conduction electron filling $\nu_c$, which we set to zero for simplicity.

Kondo physics requires that, for $\Gamma=0$, $c$-electrons form a Fermi liquid and $f$-electrons a layer of local moments formed within a large charge gap of $\approx U$, and that $\Gamma \ll U$.
We identify $U$ by computing the gap between the lower- and upper- Hubbard bands for the c-layer with a spin-polarized ground state. The use of the spin-polarized state is unphysical, but conveniently mimics, via Pauli-exclusion, the single-occupancy constraint of a featureless Mott insulator in a manner amenable to mean-field treatment without assuming a particular state of the spin-moments. Magnetic or spin-liquid ordering of the spins would give corrections to the Hubbard gap only on the order of the spin-exchange $\mathcal{O}(t_f^2/U)\ll U$, hence, the extracted value of $U$ is accurate only to this order.

We then reintroduce the interlayer tunneling $\Gamma$, which introduces (ferromagnetic) Fock-exchange coupling that competes with the Kondo exchange. However, we confirm that the latter dominates and the c/f-interaction is overall antiferromagnetic.

We note that, for the twist angle range $3.5^\circ\lesssim \theta\lesssim 6^\circ$, the SCHF occasionally predicts spin-polarized ferromagnetic states for the c-layer when $\nu_c = 1$. However, since the spin-gap for these states is much less than the antiferromagnetic Kondo coupling scale $J_K$ (which is neglected in our SCHF treatment), which would resist ferromagnetic ordering, and since SCHF often overstates the stability of charge and spin gaps, we assume that the $c$-electrons form a spin degenerate metal throughout. This assumption allows us to model the (strong) interaction effects within c-layer only as a renormalization of hopping parameters.

\begin{figure}[bt]
    \centering
    \includegraphics[width=0.9\textwidth]{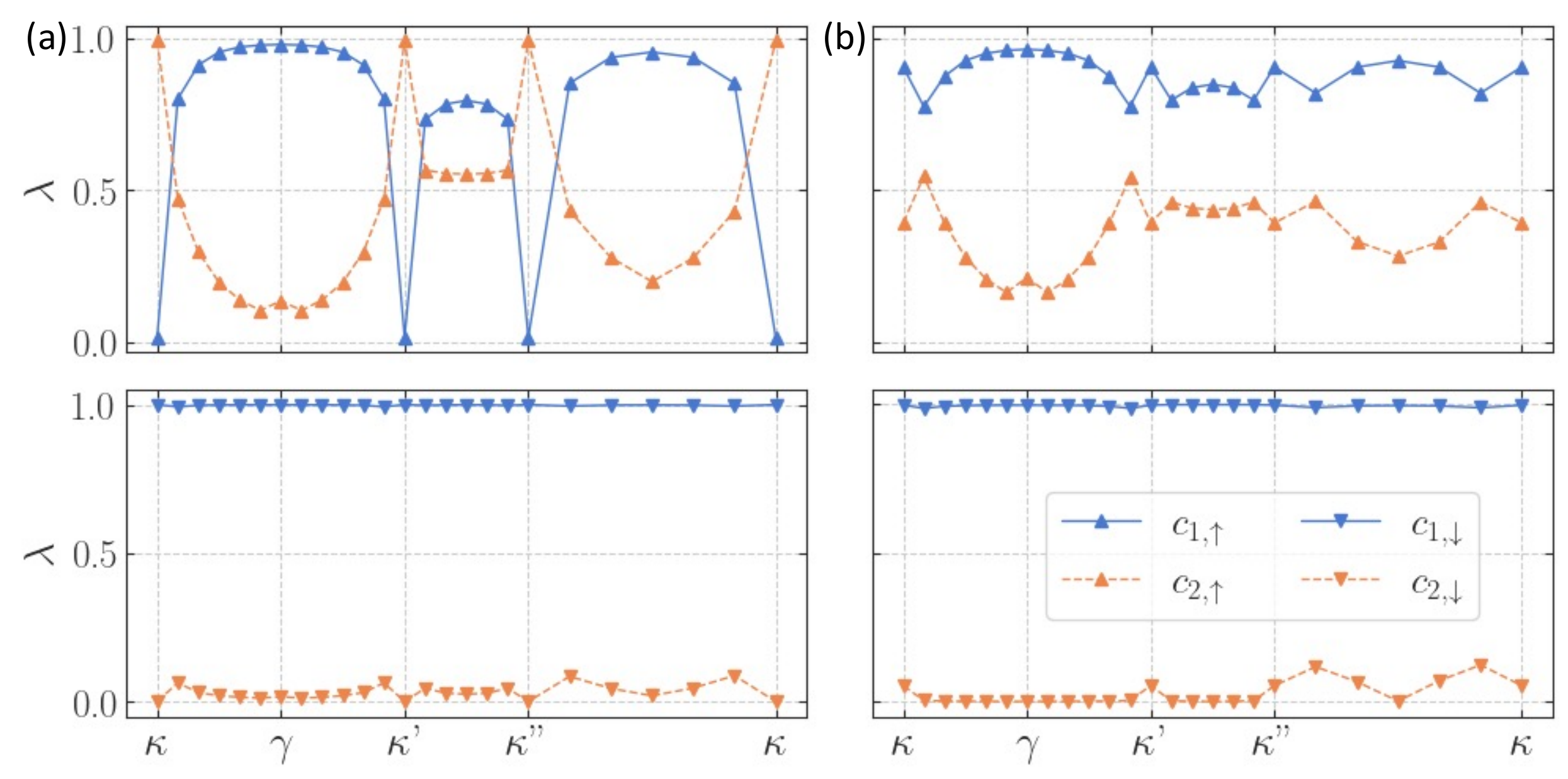}
    \caption{The tunneling suppressing factor $\lambda_{n,1}(\v{k}) = |\langle c^{HF}_{n,\sigma}|f^{HF}_{1,\sigma}\rangle|$ for $n=1$ (solid blue), $2$ (dashed orange) and $V_{mf} = 20$~meV (a), $30$~meV (b). Both panels plotted for twist angle $\theta = 4.5^{\circ}$. $|f^{HF}_{1,\uparrow/\downarrow}\rangle$ is the lower/upper Hubbard band wave function. The upper/lower panels are from the majority/minority spin.} 
    \label{fig:lambda}
\end{figure}

We model $\Gamma$ by two parameters: a $\v{k}$-independent tunneling strength parameter $w\sim 8$ meV that can be obtained from DFT calculations for $V_{mf}=V_{mc}$\cite{fengcheng2019homo} and a suppressing factor $\lambda_{n,1}(\v{k}) = |\langle c^{HF}_{n,\sigma}|f^{HF}_{1,\sigma}\rangle|$, with $n$ the band label and $|f^{HF}_{1,\uparrow}\rangle$ the lower Hubbard band, to account for the difference of the lowest-energy-band wave functions in the two layers. We make the approximation that the main contribution to the Kondo model comes from the lowest energy bands of c- and f-layer such that we can write $\Gamma(\v{k}) =3w\lambda_{1,1}(\v{k})$, where the factor of $3$ comes from the three equivalent corners of the original Brillouin zone. To understand interaction effects and justify our single-band approximation, we compare the values of $\lambda_{n,1}$ for different twist angles and $V_{mf}$s. We show the twist angle $\theta = 4.5^{\circ}$ case with $V_{mf} = 20$~meV in Fig.~\ref{fig:lambda}(a) and $V_{mf} = 30$~meV in Fig.~\ref{fig:lambda}(b) as an example. With the strong Hartree interaction (from the majority spin $f$-band) competes closely with the attractive moir\'e potential, the minority-spin $f$- and $c$-bands are close to those of free-electrons (see Fig.~\ref{fig:setup}(b) also), which explains their high values of $\lambda_{1,1}$. In particular, a tiny gap exist at $\kappa$ in the $c$-electron bands only because of avoided crossing from the moir\'e potential, which leads to a complicated overlap pattern for the majority spin in general near the zone boundary. In Fig.~\ref{fig:lambda}, we show that this complication can be avoided by increasing the moir\'e potential strength $V_{mf}$. We note, however, that there is the possibility that moir\'e potential in reality may not be strong enough for interactions. If we neglect the small regions near the Brillouin zone boundary, $\lambda_{1,1}$ is always several times larger $\lambda_{2,1}$, and shows small variance across the first Brillouin zone. For simplicity, we will henceforth assume we are in a strongly modulated regime and approximate $\lambda_{1,1}\approx 0.9$ to be $\v{k}$-independent. For direct comparison, we also tabulate the parameters for the Kondo lattice model from SCHF in Table.~\ref{tab:schf20} and Table.~\ref{tab:schf30}.

\begin{table}[hb]
    \centering
    \begin{tabular}{c|cc|ccccc}
    \hline
        Twist angle $\theta$ & $a_M$/nm & $V_{mc}$/meV & $\omega_f$/meV & $\omega_c$/meV & $t_f$/meV & $t_c$/meV & $U$/meV \\
    \hline
         $3.5^{\circ}$ & 5.40 & 7.81 & -164.37 & -0.29 & 3.94 & 6.14 & 163.60 \\
         $4.0^{\circ}$ & 4.72 & 6.82 & -182.22 & -0.16 & 5.99 & 8.13 & 177.46 \\
         $4.5^{\circ}$ & 4.20 & 5.97 & -201.34 & -0.10 & 8.32 & 10.38 & 191.78 \\
    \hline
    \end{tabular}
    \caption{Parameters from SCHF with $V_{mf} = 20$~meV.}
    \label{tab:schf20}
\end{table}
\begin{table}[h]
    \centering
    \begin{tabular}{c|cc|ccccc}
    \hline
        Twist angle $\theta$ & $a_M$/nm & $V_{mc}$/meV & $\omega_f$/meV & $\omega_c$/meV & $t_f$/meV & $t_c$/meV & $U$/meV \\
    \hline
         $3.5^{\circ}$ & 5.40 & 11.71 & -182.04 & -0.87 & 2.72 & 5.98 & 184.94 \\
         $4.0^{\circ}$ & 4.72 & 10.24 & -195.87 & -0.50 & 4.64 & 8.00 & 196.70 \\
         $4.5^{\circ}$ & 4.20 & 8.95 & -211.98 & -0.30 & 6.94 & 10.26 & 209.06 \\
    \hline
    \end{tabular}
    \caption{Parameters from SCHF with $V_{mf} = 30$~meV.}
    \label{tab:schf30}
\end{table}

We consider a dual-gated geometry with independent top gate above the c-layer bottom- gate below the f-layer, and assume that the gates are tuned such that there is one $f$-electron per \moire unit cell ($\nu_f=1$). Denote the electrical field between the active WX$_2$-layers $E_p$. Then the densities and potentials are follow from electrostatics:
\begin{align}
\begin{cases}
    \frac{4\pi e}{\epsilon}n_c &= E_p-E_t, \\
    \frac{4\pi e}{\epsilon}(n_c+n_f) &= E_b-E_t, \\
    \varphi_f-\varphi_c &= -E_pd.
\end{cases}\label{eq:em}
\end{align}

Our aim is to find a regime of gate potentials in which the itinerant electron Fermi surface lies inside the f-layer Hubbard gap. For simplicity, we neglect the small $f$-band dispersion and model the the $c$-dispersion as quadratic, with constant density of states $\rho = mA/\pi$. The energy density is therefore expressed as $\bar{\epsilon} = A\rho^{-1}n^2/2$. The classical total energy density of the system (at a fixed total density $\nu$) is then
\begin{align}
    \bar{\epsilon} =& \frac{\epsilon d}{8\pi}\(E_t+\frac{4\pi e}{\epsilon}n_c\)^2 + A\rho^{-1}n_c^2/2 + \omega_cn_c + \omega_fn_f \nonumber\\
    =& \frac{\epsilon d}{8\pi}\(D+\frac{2\pi e}{\epsilon}(n_c-n_f)\)^2 + \( A\rho^{-1}n_c^2/2 + \omega_cn_c + \omega_fn_f\),\label{eq:ebar}
\end{align}
where $\omega_c$ and $\omega_f$ are the energy minimum of the Hartree-Fock $c$-band and $f$-band respectively.
The Fermi surface condition is then $0<\mu_c-\mu_f = \partial\bar{\epsilon}/\partial n_c<U$, which simplifies to 
\begin{equation}
    -\(\frac{A\rho^{-1}}{A_M}\nu_c +\omega_c - \omega_f\)+\frac{2\pi e^2d}{\epsilon A_M}(1-\nu_c) <edD<-\(\frac{A\rho^{-1}}{A_M}\nu_c +\omega_c - \omega_f\)+\frac{2\pi e^2d}{\epsilon A_M}(1-\nu_c)+U.\label{eq:D}
\end{equation}
That is, for each $\nu_c$, it is possible to tune the displacement field to adjust the leading energy scale in the model. In the case of single-gate geometry, we see from Eq.~\eqref{eq:em} that the electric field is uniquely fixed by $n_c$ or $\nu_c$. Therefore, $D_{b/t} = \pm\frac{2\pi e}{\epsilon A_M}(1+\nu_c)$. For some parameters, it is also possible to tune through the KS-QCP utilizing the single-gate geometry for the angles we consider, consequently enabling the possibility of scanning tunneling experiments.

\section{Parton mean-field calculation}
\label{app:parton_mf}
In this section we detail the parton mean-field calculation of the phase-diagram presented in the mean text.

\subsubsection{Zero-flux states}
Expressing the spin operators in $H_\text{KH}$ in terms of spinons: $\v{S}=f^{\dagger}_a \v{\sigma}_{ab}f^{\vpd}b^{\vphantom\dagger}$, and performing standard mean-field decomposition (assuming zero net emergent magnetic flux)~\cite{senthil2004weak}, we obtain the following Hamiltonian.
\begin{align}
    H^\text{mf}_{KH} = \sum_{\v{k},\alpha\in \{\up,\down\},s} \begin{pmatrix} f^{\dagger}_{\v{k}\alpha} &c^{\dagger}_{\v{k}\alpha}
    \end{pmatrix} \begin{pmatrix}
    &\epsilon^f_{\v{k}} - \mu^f &-\gamma\\
    &-\gamma &\epsilon^c_{\v{k}} - \mu^c
    \end{pmatrix} \begin{pmatrix}
    &f^{\vpd}_{\v{k}\alpha s} \\
    &c^{\vpd}_{\v{k}\alpha s}
    \end{pmatrix} + \frac{1}{2}\sum_{\v{r},\alpha,\beta,s} \v{h}_{\v{r},s} \cdot f^{\dagger}_{\v{r}\alpha s}\v{\sigma}_{\alpha\beta}f^{\vphantom\dagger}_{\v{r}\beta s}
\end{align}
To incorporate possible $120^{\circ}$ N\'eel ordered states, we introduce a sublattice degree of freedom $s$ which runs over the three sublattices of the N\'eel pattern. Further, $\epsilon^{c}_{\v k}$, $\epsilon^{f}_{\v k}$ are the triangular lattice nearest neighbor tight-binding dispersions in terms of $t_{c}$, $\chi_f$, and $\gamma$, $\mu^f$, $\chi_f$ and $\v{h}_{\v r}$ are mean-field parameters that satisfy:
\begin{align}
    1 &= \<f^{\dagger}_{\v{r}\alpha s} f^{\vpd}_{\v{r}\alpha s}\> \nonumber\\
    2\gamma &= J_K \<c^{\dagger}_{\v{r}\alpha s} f^{\vpd}_{\v{r}\alpha s}\> \nonumber\\
    2\chi_f &= J_H \<f^{\dagger}_{\v{r}\alpha s} f^{\vpd}_{\v{r'}\alpha s}\> \nonumber\\
    4\v{h}_{s} &= J_H \sum_{\v{r'}} \<f^{\dagger}_{\v{r'}\alpha s'}\v{\sigma}_{\alpha\alpha'}f^{\vpd}_{\v{r'}\alpha' s'}\>
    \label{eq:mean_field}
\end{align}
where the sum over $\v{r'}, s'$ in the last equation runs over nearest neighbors of a given site, and there is an implicit sum over $\alpha$, $\alpha'$ and $s$. Note that we look for solutions that respect the translational symmetry of the three-sublattice pattern, and therefore, the mean-field parameters on the left hand side do not depend on the unit cell label $\v r$.

We iteratively solve Eqs.~\ref{eq:mean_field} for a fixed value of $\mu^c$, which is the chemical potential associated with the total physical charge, and $J_K$. These in turn fix $\nu$ and $D$ respectively in the phase diagram for a converged solution. At each iteration, we first obtain $\mu^f$ such that the half-filling constraint is satisfied, following which we solve the remaining mean-field equations. Note that $J_H$ is a flowing parameter in the iterative solution procedure because of its dependence on $\nu$ (which also flows along with $\chi_f$, $\gamma$ and $\v{h}_s$) via RKKY interactions. 
\begin{figure}[t]
\centering
\includegraphics[width=0.4\columnwidth]{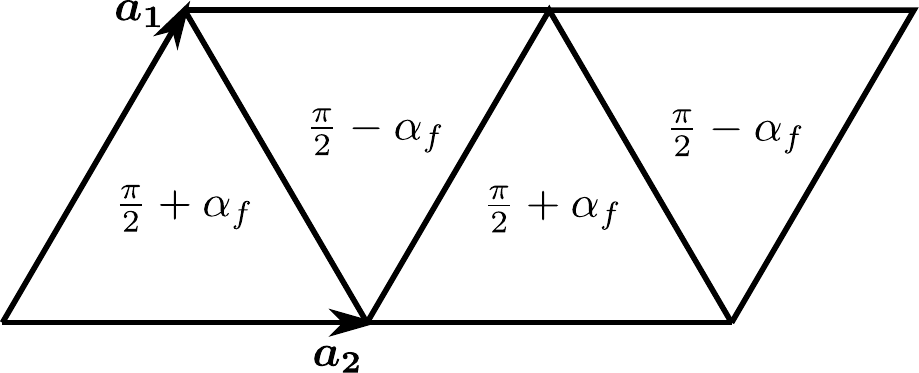}
\caption{The two-sublattice unit cell considered for the time-reversal symmetry broken states showing the background internal gauge flux through each triangle. Here $\alpha_f$ is the argument of the complex mean-field parameter $\chi_f$.}
 \label{fig_app:flux_pat} 
\end{figure}

\subsubsection{$\pi/2$-flux states}
We next generalize these mean-field solutions to incorporate non-zero flux. The analysis of~\cite{motrunich2005variational,hu2016variational} for an isolated triangular-lattice layer favors states with $\pi/2$ gauge-flux per triangle. With this in mind, we examine states with total flux $\pi$ per unit cell of the original triangular lattice (which contains a pair of triangles). We allow the flux to divide unequally between up and down-facing triangles (Fig.~\ref{fig_app:flux_pat}). This generically doubles the unit cell, as shown in Fig.~\ref{fig_app:flux_pat}, and we denote the two sublattices of the enlarged cell by $A$ and $B$ respectively. In terms of the original lattice vectors $\v{a_1}=(1/2,\sqrt{3}/2)a$ and $\v{a_2}=(1,0)a$, where $a$ is the lattice constant, we choose the new lattice vectors to be $\v{a_1}$ and $2\v{a_2}$. 

For notational clarity, let us ignore magnetic states (which we find disfavored) and drop the spin labels. The spinon Hamiltonian with the background gauge flux is then:
\begin{align}
H^\text{mf}_f = & -\chi_f e^{i\pi/6} \sum_{\v{r}}\( -f^{\dagger}_{\v{r},B} f^{\vpd}_{\v{r},A} - f^{\dagger}_{\v{r},A} f^{\vpd}_{\v{r+a_1},A} + f^{\dagger}_{\v{r+a_1},A} f^{\vpd}_{\v{r},B} +
    f^{\dagger}_{\v{r},B} f^{\vphantom\dagger}_{\v{r+a_1},B} - f^{\dagger}_{\v{r+2a_2},A} f^{\vphantom\dagger}_{\v{r},B} - f^{\dagger}_{\v{r+a_1},B} f^{\vphantom\dagger}_{\v{r+2a_2},A}\)+ \nonumber\\&+ h.c.
\end{align}
where we allow $\chi_f$ to be complex. If $\chi_f$ is real, the flux through each triangle is $\pi/2$, however for complex $\chi_f$, the flux distributes unequally between the up and the down triangles, keeping their sum fixed to $\pi$, as illustrated in Fig.~\ref{fig_app:flux_pat}. The $c$-electron part of the Hamiltonian does not contain any background gauge flux.
\begin{align}
H^\text{mf}_c =  -t_c \sum_{\v{r}}\( c^{\dagger}_{\v{r},B} c^{\vpd}_{\v{r},A} + c^{\dagger}_{\v{r},A} c^{\vphantom\dagger}_{\v{r+a_1},A} + c^{\dagger}_{\v{r+a_1},A} c^{\vphantom\dagger}_{\v{r},B} +
    c^{\dagger}_{\v{r},B} c^{\vphantom\dagger}_{\v{r+a_1},B} + c^{\dagger}_{\v{r+2a_2},A} c^{\vphantom\dagger}_{\v{r},B} + c^{\dagger}_{\v{r+a_1},B} c^{\vphantom\dagger}_{\v{r+2a_2},A}\) + h.c.
\end{align}
We also allow for a sublattice dependent $\gamma_{A/B}$, giving the following coupling between $c$ and $f$. 
\begin{align}
H^{mf}_{c-f} = -\sum_{\v{r}}\( \gamma_{A} f^{\dagger}_{\v{r},A}c^{\vpd}_{\v{r},A} + \gamma_{B} f^{\dagger}_{\v{r},B}c^{\vpd}_{\v{r},B}\)
\end{align}
The self-consistent mean-field equations are:
\begin{align}
    1 &= \<f^{\dagger}_{\v{r},A}f^{\vpd}_{\v{r},A}+f^{\dagger}_{\v{r},B}f^{\vpd}_{\v{r},B}\>\\
    \gamma_{A/B} &= J_K \<c^{\dagger}_{\v{r},A/B}f^{\vpd}_{\v{r},A/B}\>\\
    -\chi_f e^{i\pi/6} &= J_H \<f^{\dagger}_{\v{r+a_1},A}f^{\vpd}_{\v{r},A}\>
\end{align}
which we solve similar to the earlier case of zero background gauge flux. We compare the total energy of the states with and without flux at a given $\nu$ to determine the phase diagram shown in Fig.~\ref{fig:setup}. The total energy per site for the states with flux is:
\begin{align}
    \frac{E_{tot}}{N} = \frac{\<H^{mf}_c\>}{N} - \frac{|\gamma_A|^2}{2J_K} - \frac{|\gamma_B|^2}{2J_K} - \frac{3|\chi_f|^2}{J_H}
\end{align}
where $N$ is the total number of sites. The energy for the states without flux is simply obtained by using the corresponding mean-field parameters and removing the sublattice dependendence of $\gamma_{A/B}$.

\subsection{Mean-field numerical results}
\begin{figure}[t]
\centering
\includegraphics[width=\columnwidth]{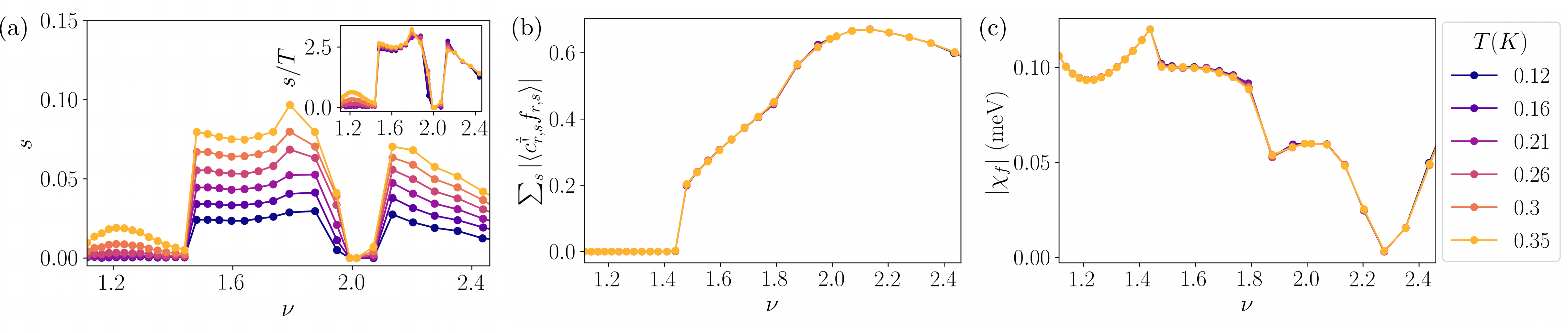}
\caption{\textbf{AH* to AH transition:} (a) Total electronic entropy per spin and per moir\'e unit cell $s$ versus conduction electron filling $\nu$ for various temperatures calculated using the parton mean-field approach for $\theta = 4.5^{\circ}$, $J_K=22.2$ meV. The inset shows $s/T$ in K$^{-1}$ versus $\nu$. (b) The hybridization order parameter (c) Spinon hopping.}
 \label{fig_app:parton_mf} 
\end{figure}

\begin{figure}[t]
\centering
\includegraphics[width=\columnwidth]{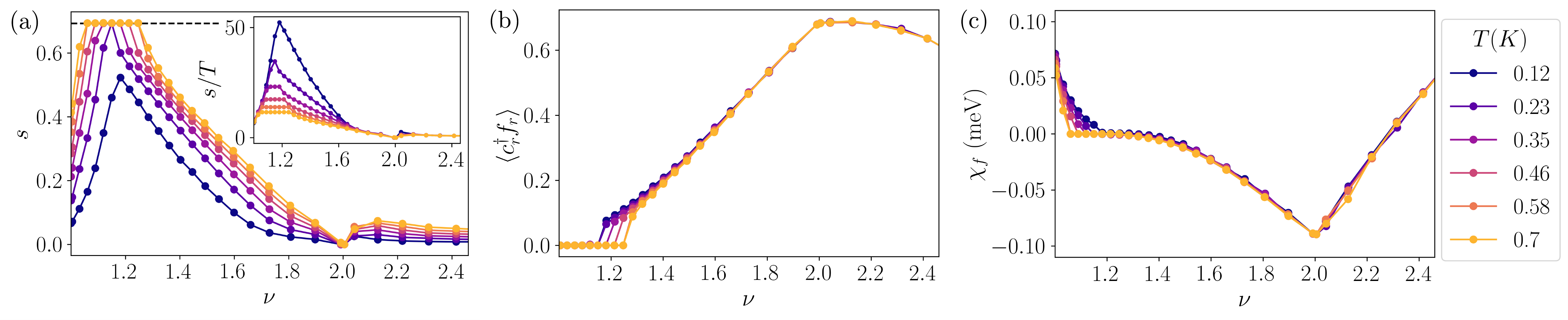}
\caption{\textbf{FL* to FL transition:} (a) Total electronic entropy per spin and per moir\'e unit cell $s$ calculated by excluding AH* and AH phases for $\theta=4^{\circ}$ and $J_K=8\Gamma^2/U=17.85$ meV. The dashed line marks $s=\log 2$. (b,c) Parton mean-field calculation results for the hybridization order parameter and spinon hopping $\chi_f$.
}
 \label{fig_app:fl_entropy}
\end{figure}

The parton mean-field prediction at finite temperatures for the finite flux and zero flux states are respectively shown in 
Fig.~\ref{fig_app:parton_mf} and Fig.~\ref{fig_app:fl_entropy}.
In the zero flux case, the KS-QCP for the FL* to FL transition is identified from the onset of the hybridization order parameter, $\Delta=\<c^\dagger f\>$, shown in Fig.~\ref{fig_app:fl_entropy}(b). At $\nu=2$ there is a KI phase. Fig.~\ref{fig_app:fl_entropy}(a) shows the calculated entropy. In the FL* phase, when the temperature is smaller than the $f$-Fermi energy, we observe regions where $s$ is linear in $T$ indicated by the approximate collapse in the $s/T$ inset plot, as expected for a Fermi liquid. This is qualitatively different from the AH* signature discussed in the main text. As the hybridization order parameter $\Delta$ onsets, $s/T$ captures the lowering of the spinon effective mass. The sign inversion of $t_f$, shown in Fig.~\ref{fig_app:fl_entropy}(c), is driven by hybridization and is not present in the single-layer parton mean-field calculation.

\subsection{Results from other twist angles}
\begin{figure*}[t]
\centering
\includegraphics[width=0.9\textwidth]{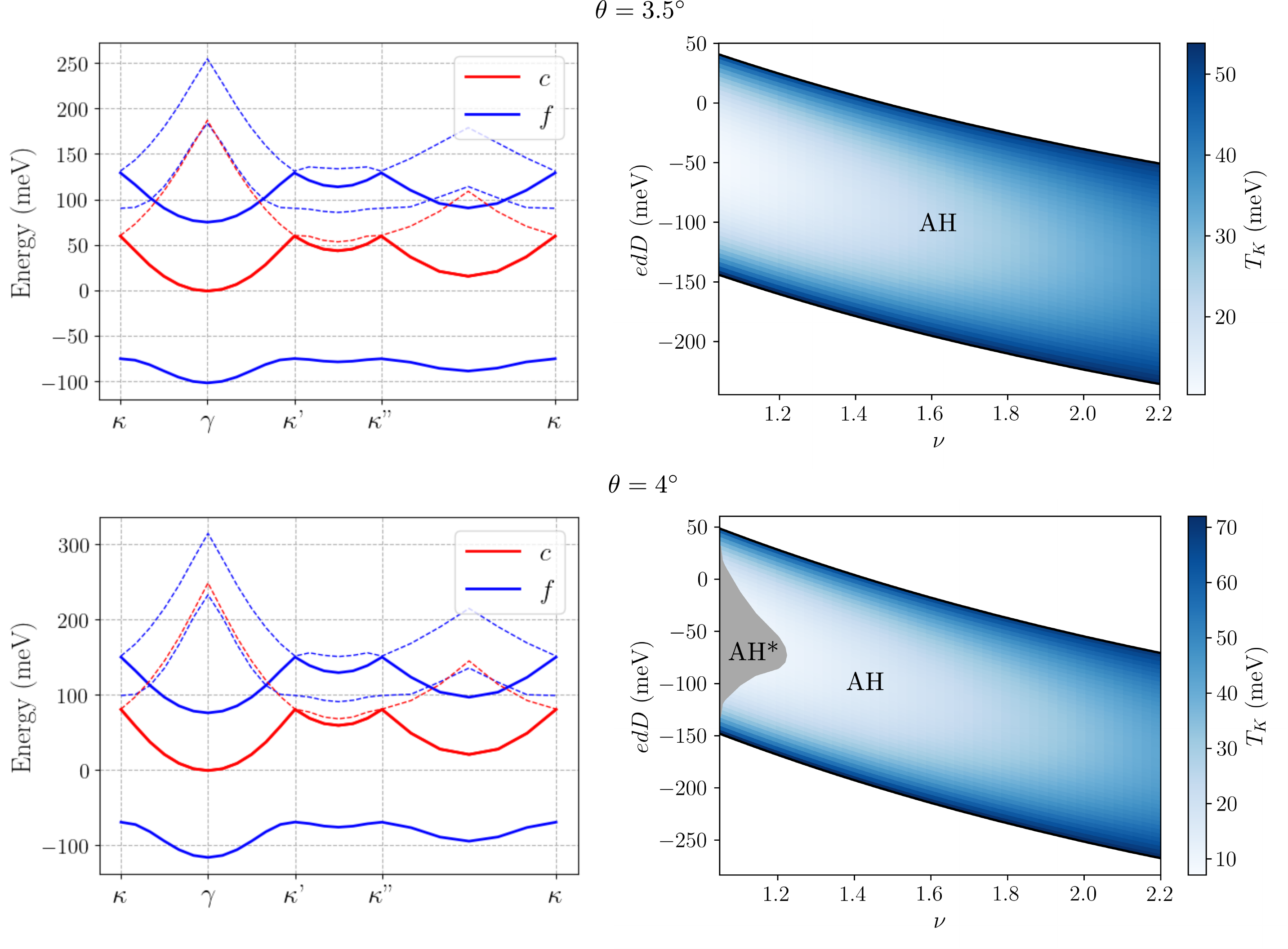}
\caption{The self-consistent Hartree-Fock band structure (left panel) and parton mean-field phase diagram (right panel) for $\theta=3.5^{\circ}$ and $\theta=4^{\circ}$. Increasing $\theta$ tends to suppress the Kondo screening and enlarge the AH* regime.} 
\label{fig_supp:other_angles} 
\end{figure*}

Fig.~\ref{fig_supp:other_angles} shows the parton mean-field phase diagram containing AH and AH* phases, for angles $\theta=3.5^{\circ}$ and $\theta=4^{\circ}$. The extracted band structure parameters are listed in Table.~\ref{tab:schf30}. $t_{c/f}$ and $U$ decrease with \moire lattice size, as expected. At $\theta=3.5^{\circ}$ the KS-QCP is absent, and the AH* region expands as $\theta$ is increased. Eventually, at even larger twist angles we expect this trend to reverse once the f-layer approaches a insulator-to-metal transition, lowering the charge gap and strengthening the tendency towards hybridization. As argued in the main text, the twist angle needs to be above a certain minimum (which is between $3.5^{\circ}$ and $4^{\circ}$ based on our calculations) to obtain a KS-QCP (for smaller angles, Kondo hybridization always dominates).

\section{RPA Transport Properties}
\label{app:ioffe_larkin}
In this section we compute the conductivity of the \moire setup by constructing effective linear response theories in the various phases in the QSL scenario.

Let us first consider the the case when the two WX\textsubscript{2} layers are simultaneously contacted, and there is a single external electromagnetic field $A$ for each of the $c,f$ layers. The effective field theory for the various phases is in terms of spinons $f$, conduction electrons $c$, the internal gauge field $a_{\mu}$ and the hybridization order parameter $\Delta$. $f$ minimally couples to $a$ as discussed in the main text, $c$ being real electrons couple minimally to $A$, and $\Delta$ couples minimally to the combination $A-a$ consistent with gauge invariance. Our objective is to obtain an effective response theory in terms of $A$. We first integrate out all the matter fields to obtain:
\begin{align}
    \L[A,a] = \frac{1}{2} A K_c A + \frac{1}{2} aK_fa + \frac{1}{2} (A-a)K_{\Delta} (A-a)
\end{align}
where the products of the form $aKa$ are shorthand for $aKa \equiv  K^{\mu \nu}(\omega,\v{q})a_{\mu}(\omega,\v{q})a_{\nu}(-\omega,-\v{q})$ where $\mu\in \{t,x,y\}$ is a space-time index, and $K_{c,f,\Delta}^{\mu\nu}(\omega,\v{q})$'s are the respective electromagnetic linear response kernels in the frequency, $\omega$, and momentum, $\v{q}$, domain. Finally integrating out $a$,
\begin{align}
    \L[A] = \frac{1}{2} A K_c A + \frac{1}{2} A\(K_{\Delta}(K_{\Delta}+K_f)^{-1}K_f \)A
    \label{eq:app_response}
\end{align}
The total response function is therefore $K_c + K_{\Delta}(K_{\Delta}+K_f)^{-1}K_f$. Considering only the spatial components, the total conductivity tensor of the system in terms of the conductivities of the constituents $f$, $\Delta$ and $c$ is given by the Ioffe-Larkin rule\cite{ioffe1989gapless}:
\begin{align}
\sigma = \sigma_c + (\sigma^{-1}_f+\sigma^{-1}_{\Delta})^{-1}
\end{align}
Let us focus on the DC transport properties across the heavy fermion quantum critical point. On the FL* side, $\Delta$ is gapped, giving a longitudinal conductivity $\sigma_{\Delta}^{xx} = 0$. On the FL side, $\Delta$ is condensed, therefore $\sigma_{\Delta}^{xx}=\infty$. At the critical point, it takes a universal value $\sigma_{\Delta}^{xx}=\sigma_0\sim \mathcal{O}(e^2/h)$~\cite{sachdev2011quantum}. Combining these behaviors with the Ioffe-Larkin rule, we obtain that the net longitudinal conductivity of the system only gets a contribution from the $c$-electrons in the FL* phase, which jumps to include the $f$ contribution in the FL phase.

We now introduce an external perpendicular magnetic field $B$ and analyze the Hall conductivity across the critical point. In order to do this, we first have to calculate the induced magnetic fields $B_{\Delta}$ and $B_{f}$ experienced by $\Delta$ and $f$ respectively. Within RPA~\cite{ioffe1989gapless,aldape2020solvable}, $B_{\Delta} = \eta B$ and $B_f = (1-\eta)B$, where $\eta$ is a ratio of magnetic susceptibilities: $\eta = \chi_f/(\chi_f+\chi_{\Delta})$. Going from FL* to FL, the magnetic susceptibility of $\Delta$ changes from $\chi_{\Delta}=\text{const.}$ to $\chi_{\Delta}=\infty$, also diverging at the critical point. In the FL* phase, since $\sigma_{\Delta}=0$, the Ioffe-Larkin rule gives a Hall conductivity $\sigma^{xy} = \sigma_{c}^{xy}(B)$. At the critical point and in the FL phase, the spinons experience the full magnetic field. Therefore the Hall conductivity jumps to $\sigma_{c}^{xy}(B)+\sigma_{f}^{xy}(B)$ in the FL phase.

\subsection{Drag Transport}
\label{app:interlayer_transport}
We also consider drag transport signatures from separately contacting the c- and f- layers. Specifically, we assume that it is possible to contact the separate layers without disrupting the Mott insulating physics of the f-layer at half-filling, so that the charge gap of the f-layer prevents a direct electrical short between contacts. While potentially challenging for fabrication, drag transport would enable one to disentangle contributions from c- and f- layers -- a capability unique to \moire settings which would be fundamentally impossible for bulk crystalline heavy fermion materials in which c- and f- electrons coexist spatially and are distinguished only by their orbital character.  
We define the dimensionless two-layer conductivity $g_{ij}(\omega) = \frac{L}{W}\frac{I_j}{\mathcal{V}_i}$ as the ratio of the current, $I_i$ in layer $i\in \{c,f\}$ to the voltage $V_j$ in layer $j$, scaled in the usual fashion with respect to the device width, $W$, and length, $L$. 

Here the c and f layers experience different external gauge fields, and we look for a response that is off-diagonal in the layer index. Let the external electromagnetic fields be $A_{c/f}$ in the $c$ and $f$ layer respectively. In addition to the hybridization order parameter, to capture (virtual) charge fluctuations above the Mott insulator of local spin moments in the f-layer, it is important to consider the fractionalized boson field $b$ that contains the charge of the local moments in the f-layer, which couples minimally to $A_f-a$. Specifically, we generalize the parton ansatz to fractionalize the electron (creation) operator in the f-layer as $\psi_\sigma = bf^{\vpd}_\sigma$, where $b$ is a spinless, charged boson. The Mott insulator of local moments at half-filling in the f-layer corresponds to a mean-field state in which $b$ form a gapped Mott insulator. This parton description is related to the Kondo lattice model considered in the main text by marginalizing over the gapped $b$ fluctuations, to obtain a description purely in terms of spinons, $f$.

Repeating the analysis of the previous subsection with obvious generalizations, we derive an effective response theory by integrating out all the matter fields:
\begin{align}
    \L[A_c,A_f,a] = \frac{1}{2} A_c K_c A_c + \frac{1}{2} aK_fa + \frac{1}{2} (A_f-a)K_b(A_f-a) + \frac{1}{2} (A_c-a)K_{\Delta} (A_c-a)
\end{align}
We now write the effective response function as a matrix in the layer index by integrating out $a$.
\begin{align}
    \L[A_c,A_f] = \frac{1}{2} \begin{pmatrix}
    A_f &A_c
    \end{pmatrix} \begin{pmatrix}
    K_b-K_b \cdot M \cdot K_b &-K_b \cdot M \cdot K_{\Delta}\\
    -K_{\Delta} \cdot M \cdot K_b &K_c+K_{\Delta} -K_{\Delta} \cdot M \cdot K_{\Delta}
    \end{pmatrix} \begin{pmatrix}
    A_f \\A_c
    \end{pmatrix}
\end{align}
where $\(K \cdot M \cdot K\)_{\mu \nu} \equiv K_{\mu \rho} M_{\rho \sigma} K_{\sigma \nu}$ and $M$ is obtained by the matrix inversion: $M \equiv \( K_f + K_b + K_{\Delta}  \)^{-1}$. The Drag conductivity is then:
\begin{align}
    g_{cf} = -\frac{\sigma_b \sigma_{\Delta}}{\sigma_f + \sigma_b + \sigma_{\Delta}}
\end{align}
Let us calculate the AC Drag conductivity. In the FL*, $\sigma_{\Delta/b} \sim -i\omega$, and therefore, Re$[g_{cf}(\omega)] \sim \frac{\omega^2}{\sigma_f}$, as discussed in the main text. In the quantum critical regime, $\sigma_{\Delta} = \sigma_0$ and $\sigma_{b} \sim -i\omega$. This gives Re$[g_{cf}(\omega)] \sim \omega^2 \sigma_0/(\sigma_f+\sigma_0)^2$. Finally in the FL phase, $\sigma_{\Delta/b} \sim 1/i\omega$ giving Re$[g_{cf}(\omega)] \sim \sigma_f/(1+\omega^2 \sigma_f^2)$.

In the AH*/AH case, introducing a Hall conductivity $\sigma_{f,H}$ and a longitudinal conductivity $\sim i\omega$ for the insulating spinons, we obtain the following results for the Hall drag conductivity.
\begin{align}
\text{Re}[g_{cf}^{xy}(\omega)] \sim
\begin{cases}
    \omega^2/\sigma_{f,H} & \text{AH*}\nonumber
    \\
    \omega^2 \sigma_0^2\sigma_{f,H}/(\sigma_{f,H}^2+\sigma_0^2)^2 & \text{KS-QCP}
    \\
    \sigma_{f,H}/(1+\omega^2
    \sigma_{f,H}^2) & \text{AH}
\end{cases}
\end{align}
Because of the hybridization between layers, electrical measurements of the $\omega=0$ drag conductivities may not be 
practical, but it may still be possible to probe finite frequency conductivities optically.  

\end{document}